\documentclass[journal]{IEEEtran}
\usepackage{amsmath}
\usepackage{cite}
\usepackage{graphicx,amssymb}
\usepackage{amsmath,amsfonts,amssymb}

\usepackage[usenames]{color}
\usepackage{algorithm}
\usepackage{algorithmic}
\usepackage{epstopdf}
\usepackage{multirow}
\usepackage{stfloats}

\begin{document}
 
\title{Adaptive $M$-QAM for Indoor Wireless Environments : Rate \& Power Adaptation}


\author{Indrakshi Dey,~\IEEEmembership{Member,~IEEE}, Geoffrey G. Messier, ~\IEEEmembership{Member,~IEEE},~and\\
Sebastian Magierowski,~\IEEEmembership{Member,~IEEE}%
\thanks{I. Dey is with CONNECT- Center for Future Networks and Communications, Trinity College Dublin, Dublin, Ireland (e-mail: deyi@tcd.ie).}
\thanks{G. G. Messier are with the Electrical and Computer Engineering, University of Calgary, Calgary, Canada (e-mail: gmessier@ucalgary.ca).}
\thanks{S. Magierowski is with the Department of Electrical Engineering and Computer Science, Lassonde School of Engineering, York University, Toronto, Canada (e-mail: magiero@cse.yorku.ca).}
}

\maketitle

\begin{abstract}

This letter presents a detailed study for indoor wireless environments, where transmit power, rate and target bit error rate (BER) are varied to increase spectral efficiency. The study is conducted for the recently proposed joint fading and two-path shadowing (JFTS) channel model, which is shown to be accurate for modeling non-Gaussian indoor WLAN environments. Analysis is done for both average and instantaneous BER constraints without channel coding, where only a discrete finite set of constellations is available. Numerical results show that, for a JFTS channel i) varying only the transmission rate (modulation constellation size) achieves more improvement in spectral efficiency compared to varying transmit power only, and ii) varying rate and/or power subject to instantaneous BER (I-BER) constraint offers better performance than when subject to average BER (A-BER) constraint. 

\end{abstract}


\IEEEpeerreviewmaketitle

\vspace*{-3mm}

\section{Introduction}\label{S1}

Over the last decade, adaptive modulation (AM) \cite{Svensson01} has emerged as a preferred technique for indoor wireless communication links (for eg. packet data CDMA standard \cite{cdma01}, IEEE 802.11$n$ \cite{ieee01}) to optimize the transmission scheme according to the state of the link. Its major advantage is that it can be designed to maintain a constant bit error rate (BER), irrespective of channel conditions, and at the same time, improving spectral efficiency. 

A plethora of channel parameters like transmit power, transmit symbol rate, modulation constellation size, instantaneous BER (I-BER) etc. can be varied to improve link spectral efficiency \cite{Steele01, Goldsmith04, Goldsmith05, Goeckel02}. Detailed studies have been presented, where one or two modulation parameters have been varied to achieve increase in spectral efficiency \cite{Steele01}, \cite{Goldsmith04}, \cite{Goldsmith05}, \cite{Cavers01, Goeckel01}, \cite{Goldsmith01}. However, how and which parameters should be adapted for maximizing spectral efficiency depends on the channel model upon which the study is based on.

Traditional fading channel models like Rayleigh, Rician or Nakagami-$m$ do not accurately characterize the indoor WLAN link, since indoor wireless links suffer from combined small-scale fading and large-scale shadowing effects. Moreover, in an indoor wireless environment, the path between the access point (AP) and the users is too short for shadowing to be accurately characterized by the log-normal distribution and the mobile users restrict their movement within a small area due to the incapability of most WLAN standards to handle soft hand-offs efficiently. Best on an extensive measurement campaign, the recently proposed joint fading and two-path shadowing (JFTS) channel model \cite{Dey01} is shown to be the best choice for characterizing such an indoor propagation scenario.

In this letter, we provide a detailed study on the increase in spectral efficiency obtained by optimally varying combinations of transmission rate, power and target BER over a JFTS faded/shadowed communication link. Numerical results demonstrate that achievable spectral efficiency over the JFTS channel enhances considerably with low target BER, an improvement much higher than exhibited by the Rayleigh fading channel \cite{Goldsmith04}. The JFTS distribution combines the Ricean fading model with the two-wave with diffused power (TWDP) \cite{Durgin01} shadowing distribution. Towards this end, to the best of the authors' knowledge, this letter is the first ever unified study on the trade-offs in adapting different AM parameters over composite faded/shadowed wireless links. Moreover, since the JFTS distribution includes a wide variety of channel conditions as special cases, this study can be readily used in many practical scenarios, both in indoor and outdoor environments. 

\vspace*{-5mm}
\section{System and Channel Model} \label{S2}

\subsection{System Model}\label{S2.1}

Let us assume a composite slow shadowed and flat faded communication channel with JFTS statistics suffering from additive white Gaussian noise (AWGN) $n[i]$ with variance $\sigma_n^2$ at time instant $i$. We transmit over this channel with average power $\overline{S}$, signal bandwidth $B$ and instantaneous signal-to-noise ratio (SNR) $\gamma[i]$. It is to be noted here that the received instantaneous SNR will be equal to $\gamma[i]$ as long as the transmit power is constant and equal to $\overline{S}$. This instantaneous SNR only reflects the influence of the channel on the SNR and not that of a varying transmit power.

For an AM technique, the instantaneous transmit power $S(\gamma[i])$ will vary depending on $\gamma[i]$. In that case, the received instantaneous SNR is equal to $\gamma[i] (S(\gamma[i]) / \overline{S})$. Since the communication link is assumed to be quasi-stationary, $\gamma[i]$ will be constant over a block of time. Hence, for simplicity of notation, we can omit the time reference $i$ relative to $\gamma$ and $S(\gamma)$. It is also noteworthy that we assume the average channel power gain is adjusted to be equal to unity through appropriate scaling of $\overline{S}$.

In this letter, we consider a family of adaptive $M$-QAM, where the choice of the available constellation sizes is restricted to $M=M_l\triangleq 2^l$ for any positive integer $l$. In this technique, the entire SNR range is divided into $L+1$ fading regions and the region boundaries are denoted by $\gamma_l$. The constellation size $M_l$ with $p_l$ bits per symbol is assigned to the $l$th fading region ($l = 0, 1, \dotso, L$). Assuming that $L$ different modulation constellations are used, the $l$th constellation $M = M_l$ is used for transmission as long as $\gamma_l \leq \gamma < \gamma_{l + 1}$ and $\gamma_L = \infty$. No signal is transmitted if $\gamma \leq \gamma_0$. We also consider that the transmit power follows the relationship, $S(\gamma) = S$ for $\gamma \geq \gamma_0$ and $S(\gamma) = 0$ for $\gamma < \gamma_0$. 

\vspace*{-5mm}

\subsection{Channel Fading Statistics}\label{S2.2}

The probability density function (PDF) of instantaneous SNR per symbol over a JFTS fading/shadowing channel can be expressed as \cite{Dey07},
\begin{align} \label{eq1}
f_{\gamma}(\gamma) =& \sum_{i = 1}^{4} \sum_{h = 1}^{m} \sum_{t = 0}^{t_{\text{max}}} \frac{A_{i, h}}{\overline{\gamma} (t!)^2} \,e^{- B_h \gamma / \overline{\gamma}} \nonumber\\
&\times \big[D_{1t} C_{1i} \big(C_{3i} \gamma / \overline{\gamma}\big)^t + D_{2t} C_{2i} \big(C_{4i} \gamma / \overline{\gamma}\big)^t\big]
\end{align}
where, $A_{i, h} = \frac{b_i R_h \Omega}{P_1 P_2} \,e^{- K - S_h}$, $B_h = \frac{\Omega}{2 P_2 r_h^2}$, $C_{1i} = \,e^{S_h \Delta T_i}$, $C_{3i} = K S_h (1 - \Delta T_i) \Omega / (P_1 P_2)$, $C_{2i} = \,e^{- S_h \Delta T_i}$, $C_{4i} = K S_h (1 + \Delta T_i) \Omega / (P_1 P_2)$, $D_{1t} = \frac{\overline{\gamma} (t!)^2}{A_{i, h}} \big(\frac{\overline{\gamma}}{C_{3i}}\big)^t \sum_{u = 1}^{t_{\text{max}} + 1} (u - 1)! \big(\frac{\overline{\gamma}}{B_h}\big)^u$, $D_{2t} = \frac{\overline{\gamma} (t!)^2}{A_{i, h}} \big(\frac{\overline{\gamma}}{C_{4i}}\big)^t \sum_{u = 1}^{t_{\text{max}} + 1} (u - 1)! \big(\frac{\overline{\gamma}}{B_h}\big)^u$, $T_i = \cos ((i - 1) \pi / 7)$, $I_0$ is the 0th-order modified Bessel function of the first kind, $m$ is the quadrature order, $R_h = \frac{w_h}{|r_h|} \,e^{r_h^2 - r_h^2 / 2 P_1}$, $\Omega$ is the mean-squared value of the JFTS envelope given by $\Omega = 4 P_1 P_2 (1 + K)(1 + S_h)$ \cite{Dey02}. In (\ref{eq1}), $\overline{\gamma}$ is the average received SNR and can be given by $\overline{\gamma} = \overline{S}/\sigma_n^2$. 

The parameter $K$ is the small scale fading parameter, $S_h$ is the shadowing parameter, $\Delta$ is the shape parameter of the shadowing distribution, $P_1$ and $P_2$ are the mean-squared voltages of the diffused and the shadowed components respectively. In (\ref{eq1}), $b_i = a_i I_0(1)$, where $a_1 = \frac{751}{17280}$, $a_2 = \frac{3577}{17280}$, $a_3 = \frac{49}{640}$ and $a_4 = \frac{2989}{17280}$. The multiplier $w_h$ denotes the Gauss-Hermite quadrature weight factors which is tabulated in \cite{Stegun01} and is given by, $w_h = (2^{m - 1} m! \sqrt{\pi}) / (m^2 [H_{m - 1}(r_h)]^2)$, where $H_{m - 1}(\cdot)$ is the Gauss-Hermite polynomial with roots $r_h$ for $h = 1, 2, \dotso, m$. 

\vspace*{-5mm}
\subsection{Instantaneous BER (I-BER)}\label{S2.3}

Assuming a square $M$-QAM with Gray-coded bits, the instantaneous BER (I-BER) as a function of $\gamma$ on an AWGN channel is approximated by, $\text{BER}(\gamma)~\approx~0.2 \,e^{- \frac{1.6 \gamma S(\gamma)}{(M_{l} -1) \overline{S}}}$, which is tight within 1 dB for $M_l \geq 4$ and $\text{BER} \leq 10^{-3}$. Hence the $\text{I-BER}$ as a function of the instantaneous SNR $\gamma$, $\text{BER}_l(\gamma_l S/\overline{S}) = \int_{\gamma_l}^{\infty} \text{BER}(\gamma) f_{\gamma}(\gamma)\mathrm{d}\gamma$ over a JFTS faded/shadowed channel can be calculated as,
\begin{align} \label{eq2}
\text{BER}_l(\gamma_l S/\overline{S}) =& \sum_{i = 1}^{4} \sum_{h = 1}^{m} \sum_{t = 0}^{t_{\text{max}}} \frac{\epsilon_{i,h,t}}{(t!)^2} \Gamma \big(t+1, \xi_h \gamma_l \big)
\end{align}
where $\epsilon_{i,h,t} = \frac{0.2~A_{i, h}(D_{1t} C_{1i} C_{3i}^t + D_{2t} C_{2i} C_{4i}^t)\overline{S}^{t+1}(M_l - 1)^{t+1}}{(B_h \overline{S}(M_l - 1) + 1.6 \overline{\gamma} S(\gamma))^{t+1}}$, $\Gamma(t+1,g) = (t!)\,e^{-g}\sum_{v=0}^{t} \frac{g^v}{v!}$ is the generalized upper incomplete Gamma function \cite{Stegun01} and $\xi_h = \frac{B_h \overline{S}(M_l - 1) + 1.6 \overline{\gamma} S(\gamma)}{\overline{S} \overline{\gamma}(M_l - 1)}$. 

\section{Rate and Power Adaptation} \label{S3}

In this section, we determine the rate region boundaries and transmit power constrained to I-BER or average BER (A-BER) for improving spectral efficiency over a JFTS channel. In particular, we study the following cases : adaptive rate and constant power with I-BER constraint (A-Rate C-Pow I-BER), adaptive rate and constant power with A-BER constraint (A-Rate C-Pow A-BER), constant rate and adaptive power with I-BER constraint (C-Rate A-Pow I-BER) and adaptive rate and power with I-BER constraint (A-Rate A-Pow I-BER). 

\vspace*{-3mm}
\subsection{A-Rate C-Pow I-BER}\label{S3.1}

We now consider the use of an I-BER constraint and a constant transmit power $S(\gamma) = S$ that is adjusted to satisfy the average power constraint, $S\int_{\gamma_{0}}^{\infty} f_{\gamma}(\gamma) \mathrm{d}\gamma = \overline{S}$ \cite{Falahati01, Goldsmith01}. It implies that the cut-off SNR $\gamma_0$ is to be chosen such that the transmit power satisfies the average power constraint. The transmit power used when transmission does occur will be higher than $\overline{S}$, and for a JFTS faded/shadowed channel can be obtained as (using integral solution from \cite[eq.~3.351.3, p.~340]{Ryzhik01}),
\begin{align} \label{eq5}
&S = \sum_{i = 1}^{4} \sum_{h = 1}^{m} \sum_{t = 0}^{t_{\text{max}}} \frac{\overline{S} A_{i, h}}{(t!)^2 B_h^{t+1}} \Gamma(t+1, B_h \gamma_{0}/\overline{\gamma}) \nonumber\\
&\qquad~~~~~~~~~~~~~~~~~~~\times (D_{1t} C_{1i} C_{3i}^t + D_{2t} C_{2i} C_{4i}^t).
\end{align}
The I-BER constraint is fulfilled at all the rate region boundaries such that, $\text{BER}_{l}(\gamma_{l}) = \text{TBER}$ where $\text{TBER}$ is the target BER. Using series expansion of Gamma function and putting it back in (\ref{eq5}), the final expression for rate region boundaries can be obtained as,
\begin{align} \label{eq6}
\gamma_l = \sum_{i = 1}^{4} \sum_{h = 1}^{m} \sum_{t = 0}^{t_{\text{max}}}\sum_{u = 0}^{t}\frac{u}{\tilde{\xi}_h} W\bigg[\hat{\xi}_h^2~\bigg(\frac{t! (u-1)!}{\hat{\epsilon}_{i,h,t}} \text{TBER}\bigg)^{1/u}\bigg]
\end{align}
where $\hat{\epsilon}_{i,h,t} = \frac{0.2~A_{i, h}(D_{1t} C_{1i} C_{3i}^t + D_{2t} C_{2i} C_{4i}^t)\overline{S}^{t+1}(M_l - 1)^{t+1}}{(B_h \overline{S}(M_l - 1) + 1.6 \overline{\gamma} S)^{t+1}}$, $\tilde{\xi}_h = \frac{B_h \overline{S}(M_l - 1) + 1.6 \overline{\gamma} S}{\overline{S} \overline{\gamma}(1 - M_l)}$, $\hat{\xi}_h = \frac{B_h \overline{S}(M_l - 1) + 1.6 \overline{\gamma} S}{\overline{S} \overline{\gamma}(M_l - 1)}$ and $W(\cdot)$ is the Product-Log function denoted as the Lambert-$W$ function tabulated in \cite{Stegun01}.

\vspace*{-3mm}
\subsection{A-Rate C-Pow A-BER}\label{S3.2}

Here, we investigate the case concerning constant power and adaptive rate under the A-BER constraint. Similar to Subsection~\ref{S3.1}, the transmit power will satisfy (\ref{eq5}). In order to maintain the A-BER constraint, the rate region boundaries should satisfy $\text{BER}_{l}(\gamma_{l}) = \text{TBER} - \frac{1}{\lambda}$ \cite{Goldsmith01}, where $\lambda \neq 0$ is the Lagrangian multiplier that satisfies the A-BER constraint. Using this constraint and following the same steps as the previous case, the final expression for rate region boundaries can be given by,
\begin{align} \label{eq7}
\gamma_l = \sum_{i = 1}^{4} \sum_{h = 1}^{m} \sum_{t = 0}^{t_{\text{max}}}\sum_{u = 0}^{t}\frac{u}{\tilde{\xi}_h} W\bigg[\hat{\xi}_h^2 \bigg(\frac{t! (u-1)!}{\hat{\epsilon}_{i,h,t}}\bigg(\text{TBER} - \frac{1}{\lambda}\bigg)\bigg)^{\frac{1}{u}}\bigg].
\end{align}
A bisection method will be used to numerically search for $\lambda$ that satisfies the A-BER constraint.

\vspace*{-3mm}
\subsection{C-Rate A-Pow I-BER}\label{S3.3}

If the transmit power $S$ is chosen such that BER becomes equal to TBER for all $\gamma \geq \gamma_0$, an I-BER constraint is fulfilled in spite of constant modulation rate. Hence,  
\begin{align} \label{eq8}
\sum_{i = 1}^{4} \sum_{h = 1}^{m} \sum_{t = 0}^{t_{\text{max}}}\sum_{u = 0}^{t} \frac{\hat{\epsilon}_{i,h,t}}{t! u!}\,e^{-\hat{\xi}_h \gamma} (\hat{\xi}_h \gamma)^u = \text{TBER}.
\end{align}
Using the expressions for $\hat{\xi}_h$, $\hat{\epsilon}_{i,h,t}$ from Subsection~\ref{S3.1} and putting them back in (\ref{eq8}), we can rewrite (\ref{eq8}) as, $\sum_{i = 1}^{4} \sum_{h = 1}^{m} \sum_{t = 0}^{t_{\text{max}}}\sum_{u = 0}^{t} (B_h \overline{S} + 1.6 \overline{\gamma} S)^{u-t-1} \,e^{-1.6 \gamma \frac{S}{\overline{S}}} = \zeta$, where $\zeta = \frac{t! u!~\text{TBER}~\overline{\gamma}^u \overline{S}^{u-t-1} \,e^{B_h \gamma / \overline{\gamma}}}{\gamma^u 0.2~A_{i, h}(D_{1t} C_{1i} C_{3i}^t + D_{2t} C_{2i} C_{4i}^t)}$. The final expression for the transmit power $S$ can be calculated as,
\begin{align} \label{eq10}
S =& \frac{\overline{S}}{1.6 \overline{\gamma}}\sum_{i = 1}^{4} \sum_{h = 1}^{m} \sum_{t = 0}^{t_{\text{max}}}\sum_{u = 0}^{t} \Bigg[(t+1-u) \nonumber\\
&\times W \Bigg[(2.71828)^{\frac{B_h \gamma}{\overline{\gamma}(t+1-u)}} \frac{\gamma \zeta^{\frac{1}{u-t-1}}}{\overline{S} \overline{\gamma} (u-t-1)}\Bigg] - B_h\Bigg]
\end{align}
It is to be noted here that the cut-off rate $\gamma_0$ must be chosen such that the average transmit power $\overline{S}$ satisfies $\int_{\gamma_{0}}^{\infty} S f_{\gamma}(\gamma) \mathrm{d}\gamma = \overline{S}$.

\vspace*{-3mm}
\subsection{A-Rate A-Pow I-BER}\label{S3.4}

In this case, both rate and power are chosen based on channel power gain information. When the I-BER is required to be equal to TBER for all SNR, the transmit power can be obtained using the constraint, $\text{BER}_{l}(\gamma_{l} S / \overline{S}) = \text{TBER}$, which finally can be expressed as,
\begin{align} \label{eq11}
S &= \frac{\overline{S}(M_l - 1)}{1.6 \overline{\gamma}}\sum_{i = 1}^{4} \sum_{h = 1}^{m} \sum_{t = 0}^{t_{\text{max}}}\sum_{u = 0}^{t} \Bigg[(t+1-u) \nonumber\\
&\times W \Bigg[(2.71828)^{\frac{B_h \gamma_l}{\overline{\gamma}(t+1-u)}} \frac{\hat{\zeta}^{\frac{1}{u-t-1}}}{\gamma_l^{t+1} \overline{\gamma} (u-t-1)}\Bigg] - B_h\Bigg]
\end{align}
where $\hat{\zeta} = \frac{t! u!~\text{TBER}~\overline{\gamma}^u \overline{S}^{u-t-1}}{0.2~A_{i, h}(D_{1t} C_{1i} C_{3i}^t + D_{2t} C_{2i} C_{4i}^t)}$. In order to obtain the optimum rate region boundaries $\gamma_l$, we need to solve \cite{Falahati01}, $S_{l-1} - S_l = \frac{p_l - p_{l-1}}{\lambda}$, where $p_{-1} = 0$ and $S_{-1} = 0$. From (\ref{eq11}), the following expression can be arrived at,
\begin{align} \label{eq12}
&\frac{\overline{S}(M_{l-1} - M_l)}{1.6 \overline{\gamma}}\sum_{i = 1}^{4} \sum_{h = 1}^{m} \sum_{t = 0}^{t_{\text{max}}}\sum_{u = 0}^{t} \Bigg[W \Bigg[(2.71828)^{\frac{B_h \gamma_l}{\overline{\gamma}(t+1-u)}}  \nonumber\\
&\times \frac{\hat{\zeta}^{\frac{1}{u-t-1}}}{\gamma_l^{t+1} \overline{\gamma} (u-t-1)}\Bigg](t+1-u) - B_h\Bigg] = \frac{p_l - p_{l-1}}{\lambda}
\end{align}
Solving the above equation, the expression for the optimum rate region boundaries $\gamma_l$ can be obtained. In this case too, a bisection method will be used to numerically search for $\lambda$ which satisfies the power constraint in (\ref{eq12}).

\vspace*{-3mm}
\section{Numerical Results and Discussion}\label{S4}

In this section, we derive analytical expressions for spectral efficiencies using the optimal rate region and power region boundaries derived for different AM techniques in Section~\ref{S3}. Next, these expressions are plotted as functions of target BERs and average received SNR ($\overline{\gamma}$) and the plots are generated by varying the fundamental parameters of the JFTS distribution. For adaptive $M$-QAM, we consider $L = 9$ transmission modes (No transmission, 2-QAM, 4-QAM, 8-QAM, 16-QAM, 32-QAM, 64-QAM, 128-QAM, 256-QAM).
\begin{figure}[t]
\vspace*{-5mm}
\begin{center}
 \includegraphics[width=1.95\linewidth]{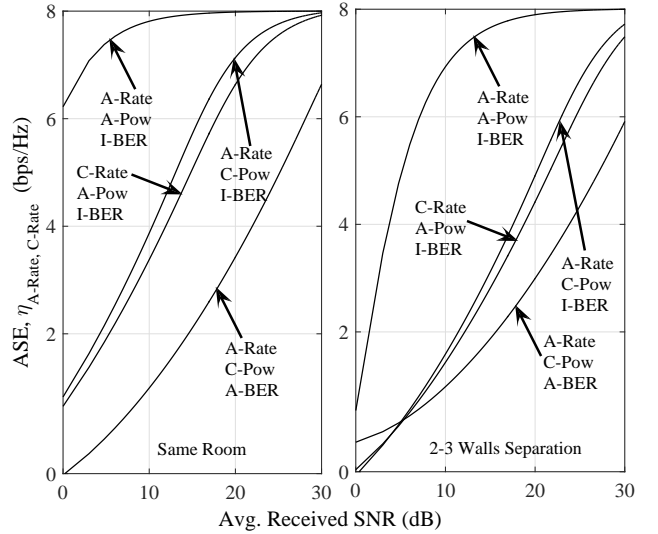}
\end{center}
\vspace*{-6mm}
\caption{Spectral Efficiency of Adaptive M-QAM at TBER of $10^{-3}$  over JFTS faded/shadowed channels when user and AP are in (a) Same Room : $K = 10$ dB, $S_h = 10.5$ dB, $\Delta =0.75$ (b) Separated by 2-3 Walls : $K = 6.5$ dB, $S_h = -1.5$ dB, $\Delta =0.25$. Plotted results are analytic only.}
\label{FIG3}
\vspace*{-6mm}
\end{figure} 
\vspace*{-5mm}
\subsection{Calculation of Average Spectral Efficiency (ASE)}\label{S4.1}

Assuming Nyquist data pulses at the lowest possible bandwidth $1/\tau_s$, where $\tau_s$ is the symbol period of the modulation, average spectral efficiency, $\eta_{A-Rate} = \sum_{l=0}^{L-1} p_{l} \int_{\gamma_{l-1}}^{\gamma_{l}} f_{\gamma}(\gamma) \mathrm{d}\gamma$, achievable over a JFTS faded/shadowed communication link can be calculated as (using integral solution from \cite[eq.~3.351.2, p.~340]{Ryzhik01}),
\begin{align} \label{eq3}
&\eta_{A-Rate} = \sum_{l=0}^{L-1} \sum_{i = 1}^{4} \sum_{h = 1}^{m} \sum_{t = 0}^{t_{\text{max}}} \frac{p_l A_{i, h}}{(t!)^2 B_h^{t+1}}(D_{1t} C_{1i} C_{3i}^t + D_{2t} C_{2i} \nonumber\\
&\quad\times C_{4i}^t) \big[\Gamma(t+1, B_h \gamma_{l}/\overline{\gamma}) - \Gamma(t+1, B_h \gamma_{l-1}/\overline{\gamma})\big].
\end{align}
For C-Rate A-Pow technique under I-BER constraint, the average spectral efficiency, $\eta_{C-Rate} =  p_{\text{max}} \int_{\gamma_{0, l}}^{\infty} f_{\gamma}(\gamma) \mathrm{d}\gamma$, with $p_{\text{max}} = \text{max} \{p_l\}$ for $0 \leq l \leq L-1$ can be calculated as,
\begin{align} \label{eq4}
\eta_{C-Rate} &= \sum_{i = 1}^{4} \sum_{h = 1}^{m} \sum_{t = 0}^{t_{\text{max}}} \frac{p_{\text{max}} A_{i, h}}{(t!)^2 B_h^{t+1}} \Gamma(t+1, B_h \gamma_{0, l}/\overline{\gamma}) \nonumber\\
&\quad \times (D_{1t} C_{1i} C_{3i}^t + D_{2t} C_{2i} C_{4i}^t).
\end{align}

\begin{figure}[b]
\vspace*{-5mm}
\begin{center}
 \includegraphics[width=1.95\linewidth]{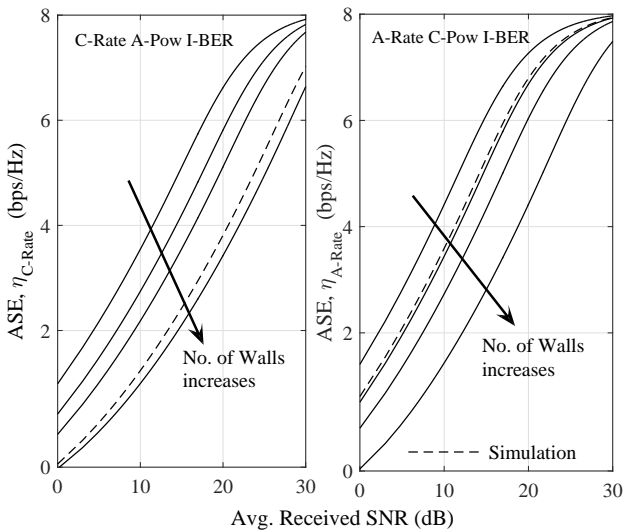}
\end{center}
\vspace*{-6mm}
\caption{Spectral Efficiency of Adaptive M-QAM at TBER of $10^{-3}$ over JFTS faded/shadowed channels with (a) A-Rate C-Pow I-BER and (b) C-Rate A-Pow I-BER. Simulation result is also plotted for comparison.}
\label{FIG2}
\vspace*{-6mm}
\end{figure} 

\vspace*{-5mm}
\subsection{Comparison of AM Techniques}\label{S4.2}

The spectral efficiency for TBER = $10^{-3}$ and two different communication scenarios (same room and 2-3 walls separation between user and AP) are illustrated in Fig.~\ref{FIG3} for the four considered techniques. The gain in the spectral efficiency under good communication link condition is considerable, as compared with poor communication scenario. Comparing different policies from the spectral efficiency point of view, we observe that for the `same room' scenario, the highest and lowest ASE are provided by A-Rate A-Pow I-BER and A-Rate C-Pow A-BER respectively due to the highest and lowest degrees of freedom respectively. However, as the link condition deteriorates, ASEs of different AM techniques approach closer to each other, except A-Rate A-Pow I-BER, which still offers considerable improvement over the other ones.

\vspace*{-3mm}
\subsection{Impact of JFTS Parameters}\label{S4.3}

Fig.~\ref{FIG2} is used to compare the effect of different JFTS parameters on the achievable ASE using AM techniques. Four different indoor WLAN communication scenarios are considered, where the user and the AP are in the same room ($K = 13$ dB, $S_h = 12$ dB, $\Delta =0.9$), separated by one ($K = 10$ dB, $S_h = 6$ dB, $\Delta =0.7$), two ($K = 7$ dB, $S_h = - 1$ dB, $\Delta =0.5$) and three ($K = 4$ dB, $S_h = - 6$ dB, $\Delta =0.3$) walls. The target BER is kept constant at $10^{-3}$ in case of both AM techniques under consideration. Achievable ASE decreases with the decrease in the JFTS parameters, $K$ and $S_h$ and gets lower than over traditional fading models \cite{Goldsmith01} as soon as $K$ decreases to 7 dB and $S_h$ to $- 1$ dB (2-walls separation scenario). The reason for this behavior is that smaller $K$ and $S_h$, poorer is the link condition with higher fading and/or shadowing severity. 

We also compare analytical results with those obtained through Monte Carlo simulation in Fig.~\ref{FIG2}. Simulated results are plotted for the case where the user and the AP are in the same room in Fig.~\ref{FIG2}(a), and for the case where the user and the AP are separated by 2 walls in Fig.~\ref{FIG2}(b). It is evident that analytical results offer good agreement with that of simulation results and they fall within 1-2 dB of the simulation results.
\begin{figure}[t]
\vspace*{-5mm}
\begin{center}
\hspace*{-10mm}
 \includegraphics[width=1.15\linewidth]{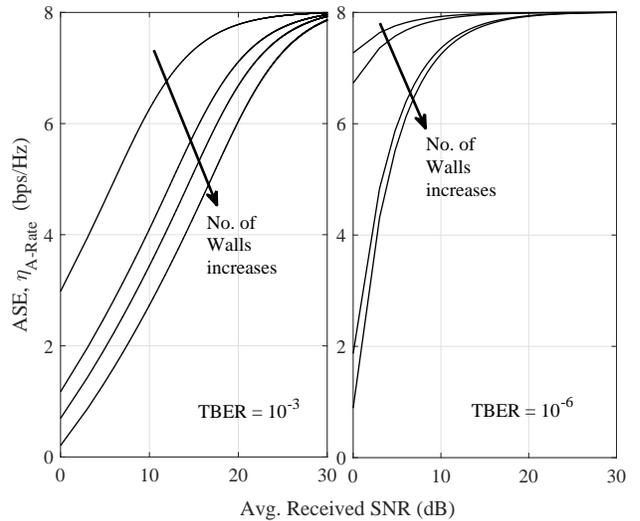}
\end{center}
\vspace*{-8mm}
\caption{Spectral Efficiency of Adaptive M-QAM with A-Rate C-Pow I-BER technique over JFTS faded/shadowed channels with two sets of TBER, (a) $10^{-3}$ and (b) $10^{-6}$. Plotted results are analytic only.}
\label{FIG1}
\vspace*{-6mm}
\end{figure} 

\vspace*{-3mm}
\subsection{Impact of Target BER}\label{S4.4}

Fig.~\ref{FIG1} demonstrates that ASE over JFTS link improves considerably with the decrease in TBER. It is worth highlighting that this improvement is much higher than over Rayleigh fading \cite{Goldsmith04}, where a huge decrease in TBER results in a very small improvement in ASE. The reason can be attributed to the fact that JFTS channel has a fading parameter $K$ always higher than 0 dB (For Rayleigh channel, $K = 0$ dB). Hence, the JFTS channel experiences less severe fading in comparison to the Rayleigh fading condition.

\vspace*{-3mm}
\section{Conclusion}\label{S5}

The main contribution of this letter is to provide a systematic study on the increase in spectral efficiency obtained by optimally varying combinations of modulation constellation size, power and target BER over a JFTS faded/shadowed indoor wireless link. Analytical results are shown to offer good agreement with that of the simulation results. Achievable ASE decreases as the number of partitions increases between the mobile user and the AP. Spectral efficiency also improves with the increase in degrees of freedom offered by different AM techniques. It is also note-worthy that ASE improves considerably with the decrease in TBER, an improvement much higher than what is observed in case of traditional fading channel models.


\begin{thebibliography}{10}

\bibitem{Svensson01}
A. Svensson, ``An introduction to adaptive QAM modulation schemes for known and predicted channels," \emph{IEEE Commun. Mag.}, vol.~95, no.~12, pp. 2322-2336, Dec. 2007.

\bibitem{cdma01}
E. T. S. Institute, ``CDMA / HDR : A bandwidth efficient high speed wireless data service for nomadic users," \emph{IEEE Commun. Mag.}, vol.~38, no.~7, pp. 70-77, Jul. 2000.

\bibitem{ieee01}
IEEE-SA, \emph{IEEE 802.11n-2009 - Amendment 5: Enhancements for Higher Throughput.} IEEE, 2009.

\bibitem{Steele01}
W. T. Webb and R. Steele, ``Variable rate QAM for mobile radio," \emph{IEEE Trans. Commun.}, vol.~COM-43, pp. 2223-2230, Jul. 1995.

\bibitem{Goldsmith04}
A. J. Goldsmith and S. G. Chua, ``Variable-rate variable-power M-QAM for fading channels," \emph{IEEE Trans. Commun.}, vol.~45, no.~10, pp. 1218-1230, Oct. 1997.

\bibitem{Goldsmith05}
A. J. Goldsmith and P. Varaiya, ``Capacity of fading channels with channel side information," \emph{IEEE Trans. Inform. Theory}, vol.~43, pp. 1218-1230, Nov. 1997.

\bibitem{Goeckel02}
C. Kse and D. L. Goeckel, ``On power adaptation in adapting signaling systems," \emph{IEEE Trans. Commun.}, vol.~48, pp. 1769-1773, Nov. 2000.

\bibitem{Cavers01}
J. K. Cavers, ``Variable rate transmission for Rayleigh fading channels," \emph{IEEE Trans. Commun.}, vol.~COM-20, pp. 15-22, Feb. 1972.

\bibitem{Goeckel01}
D. L. Goeckel, ``Adaptive coding for time-varying channels using outdated fading estimates," \emph{IEEE Trans. Commun.}, vol.~47, pp. 844-855, Jun. 1999.

\bibitem{Goldsmith01}
S. T. Chung and A. J. Goldsmith, ``Degrees of freedom in adaptive modulation: A unified view," \emph{IEEE Trans. Commun.}, vol.~49, no.~9, pp.1561-1571, Sep. 2001.

\bibitem{Dey01}
I. Dey, G. G. Messier, and S. Magierowski, ``Joint fading and shadowing model for large office indoor WLAN environments," \emph{IEEE Trans. Antennas Propagat.}, vol.~62, no.~4, pp. 2209-2222, Apr. 2014.

\bibitem{Durgin01}
G. D. Durgin, T. S. Rappaport, and D. A. deWolf, ``New analytical models and probability density functions for fading in wireless communications," \emph{IEEE Trans. Commun.}, vol. 50, no. 6, pp. 1005-1015, Jun. 2002.

\bibitem{Dey07}
I. Dey, G. G. Messier, and S. Magierowski, ``Adaptive modulation and coding for large open office indoor wireless environments," in \emph{Proc. IEEE VTC-Fall}, Sep. 2016, p. 5 pages.

\bibitem{Dey02}
I. Dey, G. G. Messier, and S. Magierowski, ``Fading statistics for the joint fading and two path shadowing channel," \emph{IEEE Wireless Commun. Lett.}, vol.~3, no.~3, pp. 301-304, Jun. 2014.

\bibitem{Stegun01}
M. Abramowitz and I. A. Stegun, \emph{Handbook of Mathematical Functions with Formulas, Graphs, and Mathematical Tables}, 9th ed. USA: New York: Dover, 1972.

\bibitem{Falahati01}
S. Falahati, A. Svensson, T. Ekman, and M. Sternad, ``Adaptive modulation systems for predicted wireless channels," \emph{IEEE Trans. Commun.}, vol.~52, no.~2, pp. 307-316, Feb. 2004.

\bibitem{Ryzhik01}
I. S. Gradshteyn and I. M. Ryzhik, \emph{Table of Integrals, Series and Products}, 7th ed. USA: Elsevier Academic Press, 2007.


\end{thebibliography}
\end{document}